\documentclass[useAMS,usenatbib]{mn2e}
\usepackage{epsfig}

\def\lsim{\raise0.3ex\hbox{$<$\kern-0.75em\raise-1.1ex\hbox{$\sim$}}}
\def\gsim{\raise0.3ex\hbox{$>$\kern-0.75em\raise-1.1ex\hbox{$\sim$}}}

\title[Observability of spheroidal galaxies]
{Observability of the virialization phase of spheroidal galaxies with radio arrays}

\author[M. Massardi et al.]{M. Massardi$^{1,2}$\thanks{E-mail: massardi@sissa.it}, A. Lapi$^{1,3}$,
 G. De Zotti$^{4,1}$, R. D. Ekers$^2$ and L. Danese$^1$ \\
$^1$ SISSA-ISAS, Via Beirut 2-4, Trieste, Italy \\
$^2$Australia Telescope National Facility, CSIRO, P.O. Box 76, Epping,NSW 1710, Australia \\
$^3$Dip. Fisica, Univ. "Tor Vergata", Via Ricerca Scientifica 1,
00133 Roma, Italy\\
$^4$INAF, Osservatorio Astronomico di Padova, Vicolo
dell'Osservatorio 5, I-35122 Padova, Italy }

\date{Accepted 2007 November 15.  Received 2007 November 13; in original form 2007 June 22}

\pagerange{\pageref{firstpage}--\pageref{lastpage}} \pubyear{2007}

\begin{document}

\maketitle
\begin{abstract}
In the standard galaxy formation scenario plasma clouds with a
high thermal energy content must exist at high redshifts since the
proto-galactic gas is shock heated to the virial temperature, and
extensive cooling, leading to efficient star formation, must await
the collapse of massive halos (as indicated by the massive body of
evidence, referred to as {\it downsizing}). Massive plasma clouds
are potentially observable through the thermal and kinetic
Sunyaev-Zel'dovich effects and their free-free emission. We find
that the detection of substantial numbers of galaxy-scale thermal
SZ signals is achievable by blind surveys with next generation
radio telescope arrays such as EVLA, ALMA and SKA. This population
is even detectable with the 10\% SKA, and wide field of view
options at high frequency on any of these arrays would greatly
increase survey speed. An analysis of confusion effects and of the
contamination by radio and dust emissions shows that the optimal
frequencies are those in the range 10--35 GHz. Predictions for the
redshift distributions of detected sources are also worked out.
\end{abstract}
\section{Introduction}\label{sec:intro}

A satisfactory theory of galaxy formation requires a good
understanding of the complex physical processes governing the
collapse of primordial density perturbations and the early galaxy
evolution. Measurements of the galaxy luminosity and stellar-mass
functions up to substantial redshifts have highlighted that these
functions show conspicuous differences with respect to the halo
mass functions predicted by the cold dark matter (CDM) theory with
the "concordance" cosmological parameters. At the low-mass end,
the halo mass functions is much steeper than the galaxy luminosity
function. As discussed by many authors (Larson 1974: Dekel \& Silk
1986; Cole 1991; White \& Frenk 1991; Lacey \& Silk 1991;
Kauffmann et al. 1993; Cole et al. 1994; Somerville \& Primack
1999; Granato et al. 2001; Benson et al. 2003), the relative
paucity of low-luminosity galaxies may be attributed to the
quenching of star formation in low-mass halos by energy injections
from supernovae and stellar winds, and by photoionization of the
pre-galactic gas. This leads to the conclusion that efficient star
formation must await the collapse of massive halos. On the other
hand, the above processes have little effect on very massive
halos, which, in the absence of additional relieving mechanisms,
would convert too large fractions of gas into stars, yielding too
many bright galaxies, with wrong metallicities  (Thomas et al.
2002; see Benson et al. 2003 and Cirasuolo et al. 2005 for
discussions of the effect of quenching mechanisms). An effective
cure for that is the feedback from active nuclei (AGNs), growing
at the galaxy centers (Granato et al. 2001, 2004; Bower et al.
2006; Croton et al. 2006)\footnote{Note that the AGN feedback
invoked by Granato et al. is radically different from that
advocated by Bower et al. and Croton et al.. The former is a
property of all AGNs and is attributed to a combination of
radiation pressure (especially line acceleration) and gas
pressure. The latter is associated to the radio active phase of
quasars (`radio mode' feedback).}.

During their very early evolutionary phases, massive
proto-galaxies are expected to contain large amounts of hot gas,
but the gas thermal history is obscure. According to the standard
scenario (Rees \& Ostriker 1977; White \& Rees 1978), the
proto-galactic gas is shock heated to the virial temperature, but
this view has been questioned  (Katz et al. 2002; Binney 2004;
Birnboim \& Dekel 2003; Kere{\v s} et al. 2005), on the basis of
independent approaches: analytic methods, a high-resolution
one-dimensional code, smoothed particle hydrodynamics simulations.
The general conclusion is that only a fraction, increasing with
halo mass, of the gas heats to the virial temperature. The hot gas
is further heated by supernova explosions and by the AGN feedback,
and may eventually be pushed out of the halo. Kere{\v s} et al.
(2005) and Dekel \& Birnboim (2006) find that there is a critical
shock heating halo mass of $\sim
10^{11.4}\hbox{--}10^{12}\,M_{\sun}$, above which most of the gas
is heated to the virial temperature, while most of the gas
accreted by less massive halos is cooler.

The large thermal energy content of the hot proto-galactic gas in
massive halos makes this crucial evolutionary phase potentially
observable by the next generation of astronomical instruments
through its free-free emission and the thermal and kinetic
Sunyaev-Zel'dovich effects (Oh 1999; Majumdar, Nath, \& Chiba
2001; Platania et al. 2002; Oh, Cooray, \& Kamionkowski 2003;
Rosa-Gonz{\'a}lez et al. 2004; De Zotti et al. 2004). In this
paper we investigate the detectability of this proto-galactic gas
exploiting an up to date model. For the purposes of the present
analysis, the adopted model can be taken as representative of the
most popular semi-analytic models (White \& Frenk 1991; Kauffmann
et al. 1993; Cole et al. 1994; Somerville \& Primack 1999; Benson
et al. 2003), that all adopt a similar mass function of dark
matter halos, a cosmological gas to dark matter ratio at
virialization, and assume that all the gas is heated to the virial
temperature.

Even if some single dish telescopes have the required theoretical
sensitivity, especially at mm and sub-mm wavelengths (e.g.
LMT/GTM, GBT at 3 mm, Rosa-Gonz{\'a}lez et al. 2004) these
continuum observations will be hampered by fluctuations in
tropospheric emission. Interferometric array observations offer a
better trade off between angular resolution, sensitivity and
control of systematics (Birkinshaw \& Lancaster 2005) together
with larger fields of view, and allow us to work at lower
frequencies where the sources of contaminations from backgrounds
and foregrounds are lower and may be better estimated. For this
reason we focus mainly on the capabilities of next generation
interferometers: the Square Kilometer Array
(SKA)\footnote{http://www.skatelescope.org/}, the Atacama Large
Millimiter Array (ALMA)\footnote{http://www.alma.info/}, the
Expanded Very Large Array
(EVLA)\footnote{http://www.aoc.nrao.edu/evla/}, the new 7 mm
capability of the Australia Telescope Compact Array (ATCA). We
note that the situation for galaxy scale SZ detection is quite
different from that for cluster SZ detection.  Cluster SZ signals
are stronger and have much larger angular scales than optimum for
the interferometer arrays and are best observed with single dishes
at high quality sites (Carlstrom, Holder \& Reese 2002).

The outline of the paper is the following: in $\S\,$\ref{sec:model}
we briefly describe our reference model; in
$\S\,$\ref{sec:counting}, we present our predictions for the counts
of proto-galaxies seen through their free-free emission and their
thermal and kinetic Sunyaev-Zel'dovich effects; in
$\S\,$\ref{sec:survey} we analyze the potential of the next
generation interferometers for detecting such signals, describe
possible survey strategies, and discuss possible contaminating
emissions and confusion effects; in $\S\,$\ref{sec:conclusions}, we
summarize our main conclusions.

We adopt a $\Lambda$-CDM cosmology with $h=0.71$, $\Omega_m=0.27$,
$\Omega_\lambda=0.73$, $\Omega_b=0.04$, $\sigma_8= 0.8$, consistent
with the results from the Wilkinson Microwave Anisotropy Probe
(WMAP) (Spergel et al. 2006).

\section{Outline of the model}\label{sec:model}

We adopt the semi-analytic model laid out in Granato et al.
(2004), with the values of the parameters revised by Lapi et al.
(2006) to satisfy the constraints set by the AGN luminosity
functions. For the reader's convenience, we summarize here its
main features.

The model is built in the framework of the standard hierarchical
clustering scenario, taking also into account the results by
Wechsler et al. (2002), and Zhao et al. (2003a; 2003b), whose
simulations have shown that the growth of a halo occurs in two
different phases: a first regime of fast accretion in which the
potential well is built up by the sudden mergers of many clumps with
comparable masses; and a second regime of slow accretion in which
mass is added in the outskirts of the halo, without affecting the
central region where the galactic structure resides. This means that
the halos harboring a massive galaxy, once created even at high
redshift, are rarely destroyed. At low redshifts they are
incorporated within groups and clusters of galaxies. Support to this
view comes from studies of the mass structure of elliptical
galaxies, which are found not to show strong signs of evolution
since redshift $z\approx 1$ (Koopmans et al. 2006). The halo
formation rate at $z \gsim 1.5$, when most massive early-type
galaxies formed (Renzini 2006), is well approximated by the positive
term in the cosmic time derivative of the cosmological mass function
(e.g., Haehnelt \& Rees 1993; Sasaki 1994).

We confine our analysis to galaxy halo masses between
$M_{\mathrm{vir}}^{\mathrm{min}}\simeq 2.5\times 10^{11}\,
M_{\sun}$, close to the mass scale at the boundary between the
blue (low mass, late type) and the red (massive, early type)
galaxy sequences (Dekel \& Birnboim 2006) and
$M_{\mathrm{vir}}^{\mathrm{max}}\approx 10^{13.2}\, M_{\sun}$, the
observational upper limit to halo masses associated to individual
galaxies (Cirasuolo et al. 2005).

The complex physics of baryons is described by a set of equations
summarized in the Appendix of Lapi et al. (2006). Briefly, the
model assumes that during or soon after the formation of the host
dark matter (DM) halo, the baryons falling into the newly created
potential well are shock-heated to the virial temperature. The hot
gas is (moderately) clumpy and cools quickly in the denser central
regions, triggering a strong burst of star formation. The
radiation drag due to starlight acts on the gas clouds, reducing
their angular momentum. As a consequence, a fraction of the cool
gas falls into a reservoir around the central supermassive black
hole (BH), and eventually accretes onto it by viscous dissipation,
powering the nuclear activity. The energy fed back to the gas by
supernova (SN) explosions and AGN activity regulates the ongoing
star formation and the BH growth. Eventually, the SN and the AGN
feedbacks unbind most of the gas from the DM potential well. The
evolution turns out to be faster in the more massive galaxies,
where both the star formation and the BH activity come to an end
on a shorter timescale, due to the QSO feedback whose kinetic
power is proportional, according to the model, to $M_{\rm
BH}^{3/2}$.

Mao et al. (2007) found that, for the masses and redshifts of
interest here, the evolution of the hot (virial temperature) gas
mass, taking into account both heating and cooling processes, is
well approximated by a simple exponential law
\begin{equation}\label{eq|Mhot}
M_{\mathrm{hot}}(t) = M_{\mathrm{hot}}(0)\,
e^{-t/t_{\mathrm{cond}}},
\end{equation}
where $M_{\mathrm{hot}}(0)=f_{\mathrm{cosm}}\, M_{\rm vir}$ is the gas
mass at virialization, $M_{\rm vir}$ being the halo mass and
$f_{\mathrm{cosm}}\approx 0.18$ the mean cosmological baryon to dark
matter mass density ratio. The evolution timescale
$t_{\mathrm{cond}}$ can be approximated as
\begin{equation}\label{eq|dtcond}
t_{\mathrm{cond}}\approx 4\times 10^8 \left(\frac {M_{\rm vir}}{10^{12}\,
M_{\sun}}\right)^{0.2}\,
\left(\frac{1+z}{7}\right)^{-1.5}~~\mathrm{yr}~.
\end{equation}
The model proved to be remarkably successful in accounting for a
broad variety of data, including epoch dependent luminosity
functions and number counts in different bands of spheroidal
galaxies and of AGNs, the local black hole mass function, metal
abundances, fundamental plane relations and relationships between
the black hole mass and properties of the host galaxies (Granato
et al. 2004; Cirasuolo et al. 2005; Silva et al. 2004, 2005; Lapi
et al. 2006).

\subsection{The virial collapse}\label{sec:virial}

The virial temperature of a uniform spherically symmetric
proto-galactic cloud with virial mass $M_{\rm vir}$ (dark matter
plus baryons) and mean molecular weight $\mu=(2X+3/4 Y)^{-1}$, X and
Y being the baryon mass fractions in the form of hydrogen and helium
(we adopt X=0.75 and Y=0.25, no metals) is
\begin{equation} \label{eq:Tvir}
T_{\rm vir}=\frac{1}{2}\frac{\mu m_p G}{k_B}\frac{M_{\rm
vir}}{R_{\rm vir}},
\end{equation}
where $m_p$ is the proton mass, G the gravitational constant, and
$k_B$ the Boltzmann constant. The virial radius $R_{\rm vir}$ is
given by
\begin{equation} \label{eq:Rvir}
R_{\rm vir}=\left(\frac{4}{3}\pi\frac{\rho_{\rm vir}}{M_{\rm
vir}}\right)^{-1/3}
\end{equation}
where $\rho_{\rm vir}$ is the mean matter density within $R_{\rm
vir}$
\begin{equation} \label{eq:Rhovir}
\rho_{\rm vir}=\rho_c  \Omega_m\Delta (1+z)^3,
\end{equation}
$\rho_c= 3H_0^2/(8\pi G)$ being the critical density. For a flat
cosmology ($\Omega_m+\Omega_\Lambda=1$), the virial overdensity
$\Delta$ can be approximated by (Bryan \& Norman 1998; Bullock et
al. 2001)
\begin{equation}
\Delta=\frac{18\pi^2+82\omega-39\omega^2}{\Omega(z)}
\end{equation}
with $\omega=\Omega(z)-1$, and
\begin{equation}
\Omega(z)=\frac{(1+z)^3 \Omega_m}{\Omega_m(1+z)^3+\Omega_\Lambda}.
\end{equation}
In the redshift range considered here ($z\ge 1.5$), we have
\begin{equation} \label{eq:Tvir2}
T_{\rm vir}\simeq 5 \times10^5 \left({M_{\rm vir} \over 10^{12}
M_{\sun}} \right)^{2/3} (1+z)\ \hbox{K},
\end{equation}
so that for the massive objects ($2.5\times 10^{11}M_{\sun}<M_{\rm
vir}<10^{13.2} M_{\sun}$) we are dealing with, the only relevant
cooling mechanism is free-free emission.

We assume that, after virialization, the protogalaxy has a NFW
density profile (Navarro, Frenk \& White 1997):
\begin{equation} \label{eq:rho}
\rho= \frac{\rho_s}{cx(1+cx)^2}
\end{equation}
where $x=r/R_{\rm vir}$,
\begin{equation} \label{eq:rho_s}
\rho_s=\frac{M_{\rm vir}}{4\pi R^3_{\rm vir}f_c}
\end{equation}
with
\begin{displaymath} \label{eq:f_c}
f_c=\frac{\log(1+c)-c/(1+c)}{c^3}
\end{displaymath}
and $c=3$ (Zhao et al. 2003b; Cirasuolo et al. 2005).

\subsection{The free-free emission}\label{sec:ff}

The free-free luminosity of the protogalaxy is computed integrating
over its volume the emissivity given by (Rybicki \& Lightman 1979):
\begin{eqnarray} \label{eq:j}
j_{\rm ff}  &\!\!\!\!=&
\!\!\!\!6.8\cdot10^{-38}n_e (\sum Z_i^2n_i) C T_{\rm vir}^{-1/2} \bar{g}_{\rm ff}(T_{\rm vir},\nu) \cdot \nonumber \\
            &\cdot& \!\!\!\!\exp(-h_P\nu/k_B T_{\rm vir})\ \hbox{erg}
\,\hbox{s}^{-1} \,\hbox{cm}^{-3}\, \hbox{Hz}^{-1},
\end{eqnarray}
where the sum in the brackets is on all the chemical species in the
gas (only H and He in our case) $Z$ being the atomic number and
$n_e$ and $n_i$ the number densities of electrons and of ions
respectively, $C$ is the clumping factor, for which we adopt the
value ($C=7$) given by Lapi et al. (2006), $h_P$ is the Planck
constant and $\bar{g}_{\rm ff}(T_{\rm vir},\nu)$ is the velocity
averaged Gaunt factor. For the latter we adopted the analytical
approximation formulae by Itoh et al. (2000) in their range of
validity. Outside such range we used the formula given by Rybicki \&
Lightman (1979):
\begin{equation} \label{eq:gff_radio}
g_{\rm ff}= \frac{\sqrt{3}}{\pi}\left[17.7+\ln\left(\frac{T_{\rm
vir}^{3/2}}{\nu}\right)\right].
\end{equation}
The gas density is assumed to be proportional to the mass density
($\rho_{\rm gas}=f_{\rm cosm}\rho$). The electron number density is
\begin{equation} \label{eq:n_e}
n_e = \frac{\rho_{gas}}{m_p}(X+Y/2),
\end{equation}
$m_p$ being the proton mass. The adopted value of the clumping
factor $C$ is assumed to be constant with radius, as in the model.
This rather crude approximation stems from our ignorance of the
complex structure of the gas distribution.

Finally, the flux scales with mass, redshift and frequency as
\begin{eqnarray} \label{eq:S_FF}
S_{\rm ff}&\!\!\!\!\!\!=& \!\!\!\!\!\!
6.6\times10^{-9}\bar{g}_{\rm ff}[T_{\rm vir},\nu(1+z)]
\left(\frac{1+z}{3}\right)^{7/2}\!\!\left(\frac{M_{\rm
vir}}{10^{12}\,M_{\sun}}\right)^{2/3}\!\!\!\!\!\!\!\!\!
\cdot \nonumber\\
    &
   \!\!\!\!\!\!\!\!\!\!\!\!\!\!\!\!\!\! \cdot &\!\!\!\!\!\!\!\!\!\!\!\!\! \left(\frac{4.8\times 10^{28}\mathrm{cm}}{d_L}\right)^2
    \!\!\exp\!\left(\frac{\!-1.9\times10^{-6}(\nu/20\,\hbox{GHz})}{
        \left(M_{\rm
        vir}/10^{12}\,M_{\sun}\right)^{2/3}}\right)\,\hbox{Jy},
\end{eqnarray}
where $d_L$ is the luminosity distance (Hogg 1999):
\begin{equation} \label{eq:dL}
d_L=\frac{c}{H_0}(1+z)\int_0^z {dz' \over
\sqrt{\Omega_m(1+z')^3+\Omega_\Lambda}}.
\end{equation}
%

\subsection{The thermal Sunyaev-Zel'dovich effects}\label{sec:sz}

The inverse Compton scattering of the Comic Microwave Background
(CMB) photons by hot electrons produces a distortion of the CMB
spectrum, known as Sunyaev-Zel'dovich (SZ) effect (Sunyaev \&
Zel'dovich 1972). The distortion consists in an increase of photon
energies which implies a decrease of the CMB brightness temperature
at low frequencies ($\nu<218\,$GHz for $T_{\rm CMB}=2.728\,$K) and
an increase at high frequencies:
\begin{equation} \label{eq:dT/T}
\frac{\Delta T_{\rm CMB}}{T_{\rm CMB}}=(x \coth{x/2}-4)y
\end{equation}
where $x=h_P \nu/(k_B T_{\rm CMB})$ and $y$ is the comptonization
parameter
\begin{equation} \label{eq:y}
y = \frac{k_B \sigma_T}{m_e c^2}\int dl\, n_e T_e
\end{equation}
$\sigma_T$ being the Thomson cross--section and $m_e$ the electron
mass. In the Rayleigh-Jeans region ($x<<1$) eq.~(\ref{eq:dT/T})
simplifies to
\begin{equation} \label{eq:dT/T_app}
\frac{\Delta T_{\rm CMB}}{T_{\rm CMB}}\simeq -2y.
\end{equation}
The SZ signal corresponds to an unresolved flux
\begin{equation} \label{eq:Ssz}
S_{tSZ}=2 \frac{(k_B T_{\rm CMB})^3}{(h_P c)^2}g(x) Y
\end{equation}
where
\begin{equation} \label{eq:gsz}
g(x)=\frac{x^4\exp(x)}{(\exp(x)-1)^2}\left(x\coth{x/2}-4\right)
\end{equation}
and $Y$ is the surface integral of the comptonization parameter.
$S_{tSZ}$ scales with mass, redshift and frequency as
\begin{eqnarray} \label{eq:S_tSZ}
S_{tSZ}&=&\!\!\!\!
0.6\times10^{-7}\left(\frac{1+z}{3}\right)^5\left(\frac{M_{\rm
vir}}{10^{12}M_{\sun}}\right)^{5/3}
\cdot \nonumber\\
    &\!\!\!\!\cdot& \!\!\!\!\left(\frac{4.8\times 10^{28}\mathrm{cm}}{d_L}\right)^2
    \left(\frac{g(x)}{0.24}\right)\,\hbox{Jy},
\end{eqnarray}
and may be positive or negative depending on the sign of $g(x)$.
Here we will quote only positive fluxes, taking the absolute value
of $g(x)$.

For a virialized cloud with $M_{\rm vir}\simeq 10^{12}\,M_{\sun}$
at $z=2$, which has a virial temperature $T_e\sim 1.4 \times
10^6\,$K, a mean electron density $n_e\simeq
10^{-3}\,\hbox{cm}^{-3}$ and a virial radius of $\simeq 106\,$kpc,
the comptonization parameter is $\simeq 10^{-7}$, yielding a
negative flux of $\simeq 60\,\hbox{nJy}$ at 20 GHz, on an angular
scale of $\simeq 10''$.

\subsection{The kinetic Sunyaev-Zel'dovich effects}\label{sec:ksz}

The kinetic Sunyaev-Zel'dovich effect is due to scattering of CMB
photons by an ionized cloud moving with peculiar velocity $v$. The
associated flux density is (Carlstrom, Holder \& Reese 2002):
\begin{equation} \label{eq:Sszk}
S_{kSZ}=-\frac{v_{p}}{c}\ 2 \frac{(k_B T_{\rm CMB})^3}{(h_P
c)^2}h(x) \int{\tau_e d\Omega}
\end{equation}
where
\begin{equation} \label{eq:hsz}
h(x)= \frac{x^4\exp(x)}{(\exp(x)-1)^2},
\end{equation}
$\tau_e$ is the optical depth, $v_{p}$ is the line-of-sight
component of the velocity, and the surface integral of $\tau_e$ is
carried out over the solid angle of the moving cloud. The resulting
CMB spectrum is still Planckian, but shifted towards higher (lower)
temperatures for negative (positive) velocities (where negative
means towards the observer). The function $h(x)$ has a maximum at
$\nu=218$ GHz where the thermal effect vanishes. $S_{kSZ}$ scales as
\begin{eqnarray} \label{eq:S_kSZ}
S_{kSZ}&=& 5.3\times10^{-8}\left({|v|/393\,
\mathrm{km/s}}\right)\left[{(1+z)/3}\right]^4
\cdot \nonumber\\
    &\!\!\!\!\!\!\!\!\!\!\!\!\!\!\!\!\!\!\!\!\!\!\!\!\!\!\!\!\!\!\!\!\!\!\!\!\!\cdot&
  \!\!\!\!\!\!\!\!\!\!\!\!\!\!\!\!\!\!\!\!\!\!\!\!\!\!\! \left({M_{\rm
vir}/10^{12}M_{\sun}}\right)  \left({4.8\times
10^{28}\mathrm{cm}/d_L}\right)^2
    \left[{h(x)/0.12}\right]\,\hbox{Jy}.
\end{eqnarray}
Following Sheth \& Diaferio (2001) we model the distribution
function of galaxy peculiar velocities, $P(v)$, as a Gaussian core
with $\sigma_v=680 (1+z)^{-1/2}\,\hbox{km}\,\hbox{s}^{-1}$,
extending up to $v_t=1742 (1+z)^{-1/2}\,\hbox{km}\,\hbox{s}^{-1}$,
followed by exponential wings cut off at $v_{\rm max}=3000
(1+z)^{-1/2}\,\hbox{km}\,\hbox{s}^{-1}$. Normalizing the integral
of $P(v)$ to unity, we have:
\begin{eqnarray} \label{eq:Pv}
P(v)dv &\!\!\!\!=&\!\!\!\! 5.86 \times 10^{-4}(1+z)^{1/2} \,(dv/\hbox{km}\,\hbox{s}^{-1})\cdot \nonumber\\
    &\!\!\!\!\!\!\!\!\!\!\!\!\!\!\!\!\cdot& \!\!\!\!\!\!\!\!\!\!\!\!\left\{\begin{array}{ll}
            \exp[-0.5(v/\sigma_v)^2] & \mathrm{for}~|v|\le v_t \\
            2.065 \exp(-2.3 |v|/v_0) & \mathrm{for}~ v_t <  |v| \le v_{\rm max}
\end{array}
\right.
\end{eqnarray}
where $v_0=1000 (1+z)^{-1/2}\,\hbox{km}\,\hbox{s}^{-1}$. The adopted
scaling with redshift is that appropriate in the linear regime, when
the effect of the cosmological constant can be neglected, as is the
case in the $z$ range of interest here.

\begin{figure}
\begin{center}
\includegraphics[width=6cm, angle=90]{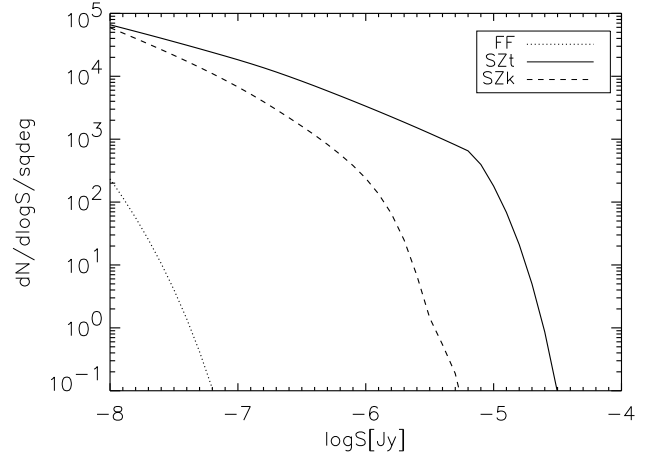}
\end{center}
\caption{Comparison of the differential source counts at 20 GHz of
thermal (solid lines) and kinetic (dashed lines)
Sunyaev-Zel'dovich effects and free-free (dotted line). For SZ
effects we obviously use the absolute value of the flux. The
counts of the kinetic SZ effect include both positive and negative
signals, and are therefore a factor of 2 larger than those given
by eq.~(\protect\ref{eq:dNdlogS_kin}).  The decline of the counts
of Sunyaev-Zel'dovich effects at the faint end is due to the
adopted lower redshift ($z\ge 1.5$) and halo mass ($M_{\rm vir}\ge
2.5\times 10^{11} M_{\sun}$) limits. } \label{fig:comp_all}
\end{figure}

\begin{figure}
\begin{center}
\includegraphics[width=6cm, angle=90]{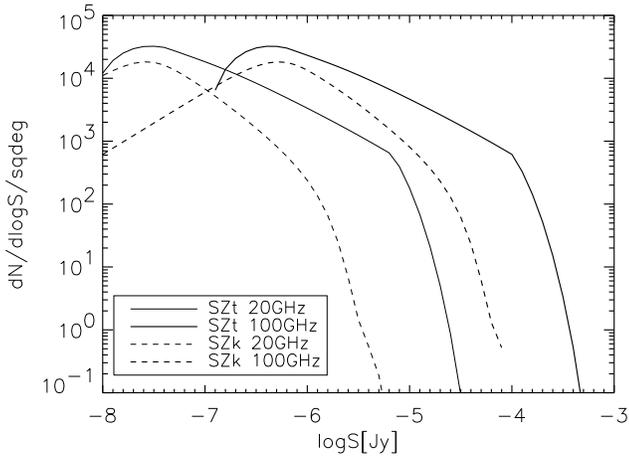}
\end{center}
\caption{Comparison of the differential source counts at 20 GHz
and 100 GHz of thermal (thin and thick solid lines respectively)
and kinetic (thin and thick dashed lines) Sunyaev-Zel'dovich
effects. } \label{fig:comp_SZ}
\end{figure}

\section{Source counts}\label{sec:counting}

The mean differential number counts per steradian are given by:
\begin{equation} \label{eq:dNdlogS_gen}
\frac{d\cal{N}(S)}{d\log S}=\int_{\ln(z_0)}^{\ln(z_1)} d\ln(z)\, z
\frac{dV}{dz}n[L(S,z),z]{d\log L\over d\log S}
\end{equation}
where $n(L,z)$ is the comoving epoch-dependent luminosity function
per unit $d\log L$, $dV/dz$ the comoving volume per unit solid
angle:
\begin{equation} \label{eq:dVdz}
\frac{dV}{dz}=\frac{c}{H_0(1+z)^2}\frac{d_L^2}{\sqrt{\Omega_m(1+z)^3+\Omega_\Lambda}}.
\end{equation}
According to our model, for any given $z$ the free-free luminosity
of a proto-spheroid and its thermal SZ signal depend only on its
virial mass. The luminosity function can then be straightforwardly
computed integrating the formation rate of virialized objects,
$d^2N(M_{\rm vir},z)/d\log M_{\rm vir}dt$, over the duration of
the ionized phase and multiplying the result by $d\log M_{\rm
vir}/d\log L$. To avoid unnecessary complications we keep the
free-free luminosity constant at its initial value over a time,
$t_{\rm ion}$, equal to the minimum between $t_{\mathrm{cond}}$
[eq. (\ref{eq|dtcond})], the expulsion time of the interstellar
gas, $\Delta t_{\mathrm{burst}}$, determining the end of the star
formation burst, and the expansion timescale, and zero afterwards.
This simplifying assumption implies that the evolution of the hot
gas mass [eq.~(\ref{eq|Mhot})], density, clumping factor, and
temperature are neglected. It is motivated by our expectation that
the effect on the free-free luminosity, hence on the counts, of
the moderate decrease of the hot gas mass over the time $t_{\rm
ion}$ is counterbalanced by an increase of the mean gas density
and of the clumping factor, as a consequence of the shocks
associated to supernova explosions and to the AGN feedback. Also,
having neglected the contribution to the counts from the free-free
emission at $t > t_{\rm ion}$, partly compensates the possible
overestimate due to having neglected the decrease of the gas mass.
In any case, a more sophisticated calculation does not appear to
be warranted since, as discussed in \S\,\ref{sec:contaminants} and
\ref{sec:confusion}, the free-free signal turns out to be too weak
to be detectable, being overwhelmed by emissions associated to
star formation.

Mao et al. (2007) found that $\Delta t_{\mathrm{burst}}$ (yr) can be approximated as
\begin{eqnarray}\label{eq|dtburst}
\Delta t_{\mathrm{burst}}\!\!\!\!\!&\approx&\!\! \!\!\!4\times 10^8
\, \left({1+z\over
7}\right)^{-1.5} \cdot \nonumber \\
\!\!\!\!\!\!\!\!\!&\cdot&\!\!\!\!\! \left\{\begin{array}{ll}\!\! 1
 & \!\!\mathrm{for}\, M_{\rm vir}\ge 10^{12} M_{\sun} \\
\!\!\left(M_{\rm vir}/10^{12} M_{\sun}\right)^{-0.15} &
\!\!\mathrm{for}\, M_{\rm vir}< 10^{12} M_{\sun}
  \end{array}
\right. \!\!\!\!\! .
\end{eqnarray}
The mass function of ionized protospheroids at the redshift $z$ is
then:
\begin{equation} \label{eq:MFion}
\left({dN_{\rm ion}(M_{\rm vir},z)\over dM_{\rm vir}}\right)_{\rm
ion}=\int_{t(z)-t_{\rm ion}}^{t(z)}dt'{d^2N(M_{\rm vir},z)\over
dM_{\rm vir}\, dt'}.
\end{equation}
The formation rate of protospheroids is well approximated by the
positive term of the derivative of the Sheth \& Tormen (1999) mass
function, $(dN(M_{\rm vir},z)/dM_{\rm vir})_{\rm ST}$, (Lapi et al.
2006):
\begin{eqnarray} \label{eq:dNdMdt}
\frac{d^2N(M_{\rm vir}, z)}{dM_{\rm vir} dt}\!\!\!\!\!&=&\!\!\!\!\!
\left[\frac{a \delta_c(z)}{\sigma^2(M_{\rm vir})}+
\frac{2p}{\delta_c(z)} \frac{\sigma^{2p}(M_{\rm
vir})}{\sigma^{2p}(M_{\rm vir})+a^p
\delta_c^{2p}(z)}\right] \cdot \nonumber \\
\!\!\!\!\!&\cdot& \!\!\!\!\!\left({dN(M_{\rm vir},z)\over dM_{\rm
vir}}\right)_{\rm ST} \left|\frac{d\delta_c(z)}{dt}\right|
\end{eqnarray}
where $a=0.707$, $p=0.3$, $\delta_c(z)$ is the critical overdensity
for the spherical collapse, $\sigma(M_{\rm vir})$ is the rms
amplitude of initial density fluctuations smoothed on a scale
containing a mass $M_{\rm vir}$. In turn, the Sheth \& Tormen (1999)
mass function writes
\begin{equation}
\left({dN(M_{\rm vir},z)\over dM_{\rm vir}}\right)_{\rm ST}=\frac
{\rho}{M_{\rm vir}^2} \nu f(\nu) \frac {d\ln \nu } {d \ln M_{\rm
vir}}
\end{equation}
where $\rho$ is the average comoving density of the universe,
$\nu=[\delta_c(z)/\sigma_\delta(M_{\rm vir})]^2$, and
\begin{equation}
\nu f(\nu) =A[1+(a\nu)^{-p}]\left(\frac
{a\nu}{2}\right)^{1/2}\frac{e^{-a\nu /2}}{\pi ^{1/2}},
\end{equation}
with $A=0.322$.

The calculations leading to the counts of the thermal SZ ``fluxes''
are strictly analogous. In the case of the kinetic SZ effect we
need also to take into account the redshift dependent distribution
of peculiar velocities, and we have
\begin{eqnarray} \label{eq:dNdlogS_kin}
\frac{d\cal{N}(S_{\rm kSZ})}{d\log S_{\rm kSZ}}&=&\int_{\ln(z_0)}^{\ln(z_1)} d\ln(z)\, z \frac{dV}{dz} \int_{\ln(v_{\rm min})}^{\ln(v_{\rm max})}\!\!\!\!\!\!\!\!\!\!\! d\ln(v)\, v P(v) \cdot \nonumber \\
\!\!\!\!\!\!\!\!\!\!\!&\cdot&\frac{dN_{\rm ion}[M_{\rm vir}(z, v)]}{d\log M_{\rm vir}}{d\log M_{\rm vir}\over d\log S_{\rm kSZ}},
\end{eqnarray}
where $v_{\rm min}$ is the velocity yielding a kinetic SZ ``flux''
$S_{\rm kSZ}$ from a galaxy with the maximum considered mass
($M_{\rm vir}=10^{13.2} M_{\sun}$) at redshift $z$, $dN_{\rm
ion}(M_{\rm vir}, z, v)/dM_{\rm vir}$ is the differential mass
function of proto-spheroidal galaxies with peculiar velocity $v$
and redshift $z$, producing a kinetic SZ flux $S_{\rm kSZ}$. As
before, $dN_{\rm ion}(M_{\rm vir}, z, v)/dM_{\rm vir}$ is computed
integrating the formation rate of virialized objects over the
duration of the ionized phase. Equation~(\ref{eq:dNdlogS_gen})
gives the number of either positive or negative kinetic SZ
signals. The comparison of the differential source counts at 20
GHz in Fig. \ref{fig:comp_all} shows that the thermal
Sunyaev-Zel'dovich effect is dominant above $10^{-8}\,$Jy. The
decline of the SZ counts at faint flux levels is due to the
adopted lower limits to halo masses and redshifts ($M_{\rm vir}\ge
2.5\times 10^{11} M_{\sun}$ and $z\ge 1.5$). The very steep slope
at the bright end comes from the high halo mass cutoff. The
free-free counts are very low, indicating that this emission is
very hard to detect in the radio.

As illustrated by Fig.~\ref{fig:comp_SZ}, the SZ fluxes increase
with increasing frequency in the Rayleigh-Jeans region of the Cosmic
Microwave Background.

\section{Perspectives for searches of ionized proto-spheroidal clouds}\label{sec:survey}
\subsection{Next generation mm-wave interferometers}
In Table \ref{tab:interferometers} we have collected some of the
main properties of next generation radio interferometers working
at few cm to mm wavelengths.

\begin{table}
\caption{Main properties of next generation interferometers. The
maximum baseline has been calculated considering that the angular
size, for the galaxies in the intervals of mass and redshift we
are considering, ranges from $5''$ to $35''$, and requiring a
ratio of 5 between amplitude and noise on the visibilities. 10\%
SKA has the same properties as SKA, but the number of baselines is
$1.25\times 10^5$} \label{tab:interferometers}
\begin{tabular}{|l|c|c|c|c|}
\hline
                    & FULL-SKA          &  ALMA & ATCA   & EVLA \\
 \hline
Frequency (GHz)     & 10-20             & 100   & 35-50  & 35   \\
Bandwidth(GHz)      & 4                 & 4x2   & 2x2    & 8    \\
Antenna diam. (m)   & 12 m              & 12    & 22     & 25   \\
Efficiency          & 0.8               & 0.8   & 0.8    & 0.8  \\
$T_{\rm sys}$ (K)   & 50                & 50    & 60-80  & 75   \\
No. of polariz.     & 2                 & 2     & 2      & 2    \\
Min. baseline (m)   & 15                & 15    & 30.6   & 30   \\
Max. baseline (km)  & 1.4-0.7           & 0.2   & 0.4-0.3& 0.4  \\
No. of baselines    & $1.25\times 10^7$ & 700   & 10     & 350  \\

\hline
\end{tabular}
\end{table}

The Australia Telescope Compact Array (ATCA) is a 6 22m-dish array.
The technical parameters we use here refer to the recently completed
upgrade to the 7 mm receivers and the increase of the bandwidth from
the present $2\times128 \mathrm{MHz}$ up to 4 GHz (CABBS). The band
ranges between 30 to 50 GHz, with $T_{\rm sys}$ increasing from 60 K up
to 80 K at the top end of the band. The system will be fully
operational by 2008.

The Atacama Large Millimiter Array (ALMA) is a 50 12 m antenna
array. The lower frequency band with higher priority ranges between
84 and 116 GHz, close to the maximum amplitude in flux of the
negative signal of thermal Sunyaev-Zel'dovich effect. The array will
be operational by 2012.

The Expanded Very Large Array (EVLA) is an improvement of the
sensitivity, frequency coverage, and resolution of the existing
VLA. When completed, after 2013, it will use the 27 25m dishes of
VLA working in the frequency range 1-50 GHz with 8 GHz bandwidth
per polarisation available in the frequency bands 18-26.5,
26.5-40, and 40-50 GHz.

The Square Kilometer Array (SKA) is a titanic project for an
interferometer whose main technical specification is to have at
least one square kilometer of detecting area in the core region.
The highest frequency band should span the range 16--25 GHz.
Several designs are under consideration. The parameters we use
refer to the small parabolic dishes version, which is the only
high frequency design being considered. The telescope is expected
to be fully operational after 2020, but a ``10\% SKA'' is expected
to be operating as early as 2015. Phased array feeds in the focal
plane are being considered for the lower frequency receivers. If
such systems were implemented at the higher frequencies they would
increase the field of view and hence the survey speed by factors
of up to 50.

The angular resolution of an array of antennas is given by
\begin{equation}\label{eq:resolution}
 \theta = 1.02 \frac{\lambda}{B}
\end{equation}
where $B$ is the maximum distance between two antennas. The field
of view normally corresponds to the Half Power Beam Width (HPBW)
of an antenna
\begin{equation}\label{eq:HPBW}
 \hbox{HPBW}= 1.02 \frac{\lambda}{D}
\end{equation}
where $D$ is the diameter of the antenna dish. For a Gaussian beam
the field of view (FOV) is
\begin{equation}\label{eq:FOV}
 \hbox{FOV}= {\pi\over \ln\,2 }\left({\hbox{HPBW}\over 2}\right)^2.
\end{equation}
Phased array feedhorns add a multiplying factor to this relation,
increasing by the same factor the sky area covered in a single
pointing.
The noise level in an image is given by
\begin{equation}\label{eq:sensitivity}
 \sigma_{\rm image}=\frac{k_B T_{\rm sys}}{A \eta}\sqrt{\frac{1}{t\  N_{\rm base}\Delta\nu \ n_{\rm pol}}}.
\end{equation}
where $T_{\rm sys}$ is the system temperature, $A$ is the antenna
surface area, $\eta$ is the system (dish and receiver) efficiency,
$t$ is the integration time, $N_{\rm base}$ is the number of
baselines short enough to have full sensitivity to observe objects
with size between 5'' and 35'', $\Delta\nu$ is the bandwidth and
$n_{\rm pol}$ is the number of polarizations. Considering the NFW
profile for densities enlarges the range of full sensitivity
baselines in the visibility space improving the resolution without
losing too much in sensitivity. We made our calculations using a
reasonably conservative configuration.

Reference values of the quantities used in the calculations for
the instruments mentioned above are given in
Table~\ref{tab:interferometers}.

\begin{figure}
\begin{center}
\includegraphics[width=6cm, angle=90]{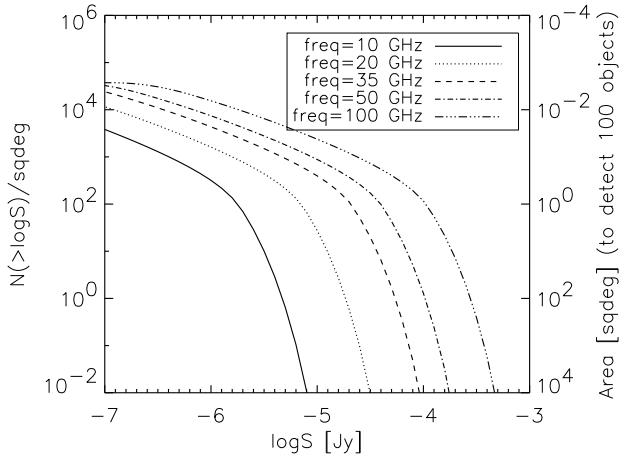}
\end{center}
\caption{Integral counts and sky area required to detect the thermal
SZ effect of 100 protospheroids (right-hand scale) as a function of the absolute value of the ``flux'' at 20, 35 and 100 GHz. }
\label{fig:intcount}
\end{figure}

\begin{figure}
\begin{center}
\includegraphics[width=6cm, angle=90]{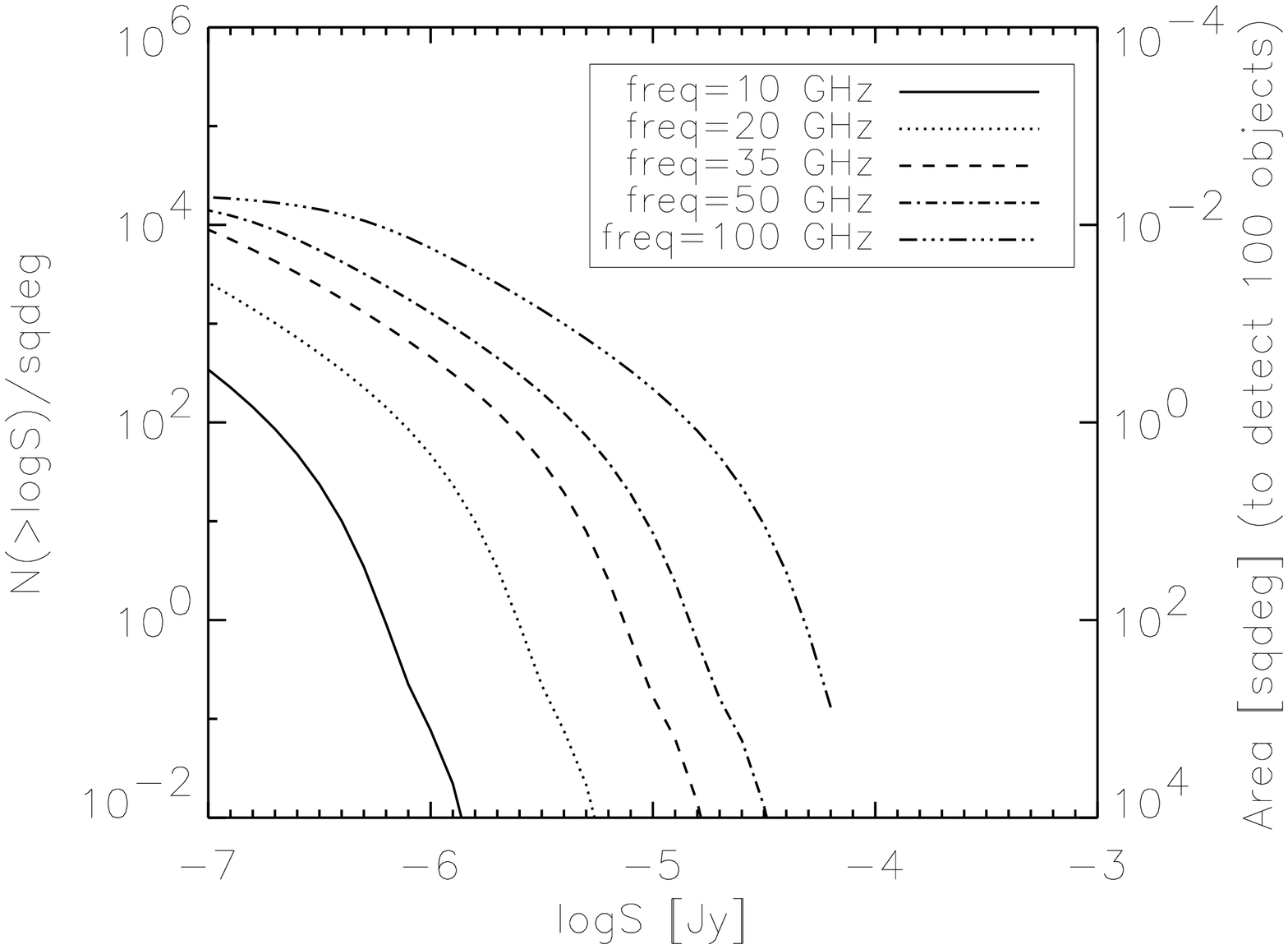}
\end{center}
\caption{Same as Fig.~\ref{fig:intcount} but for the kinetic SZ
effect. As in Fig.~\ref{fig:comp_all}, the counts include both
positive and negative signals; for the latter, $S$ is obviously
the absolute value of the flux.} \label{fig:intcount_SZk}
\end{figure}

Assuming as detection level of an object emitting flux $S_{\rm lim}$
the ratio $S_{\rm lim}/\sigma_{\rm image}=5$, for a given telescope
the integration time required for each pointing can be obtained by
inverting eq.~(\ref{eq:sensitivity}).  The number of pointings
necessary to cover a sky area $A_s$ is
\begin{equation}\label{eq:num_point}
n_p=A_s/FOV.
\end{equation}
If the integral counts of sources scale as $S^{-\beta}$, the
number of sources detected in a given area scales as
$t^{\beta/2}$. For a given flux, the number of detections is
proportional to the surveyed area, i.e. to $t$.  Thus, to maximize
the number of detections in a given observing time we need to go
deeper if $\beta>2$ and to survey a larger area if $\beta<2$. The
number of sources detected above a given flux limit, $S_{\rm
lim}$, within a telescope FOV, $N_{\rm FOV}$, is straightforwardly
derived from the source counts. The number of such pointings
necessary to detect $N_{\rm s}$ sources is $n_{\rm p} = N_{\rm
s}/N_{\rm FOV}(S_{\rm lim})$ and the corresponding surveyed area
is $A_{\rm s} = n_{\rm p}\,\hbox{FOV}$. The predicted integral
counts of thermal and kinetic SZ effect for several frequencies,
covered by the radio interferometers mentioned above, are shown in
Figs.~\ref{fig:intcount} and \ref{fig:intcount_SZk}, respectively.
The scale on the right-hand side of these figures gives the
corresponding area containing 100 protospheroids.

\begin{figure}
\begin{center}
\includegraphics[width=6cm, angle=90]{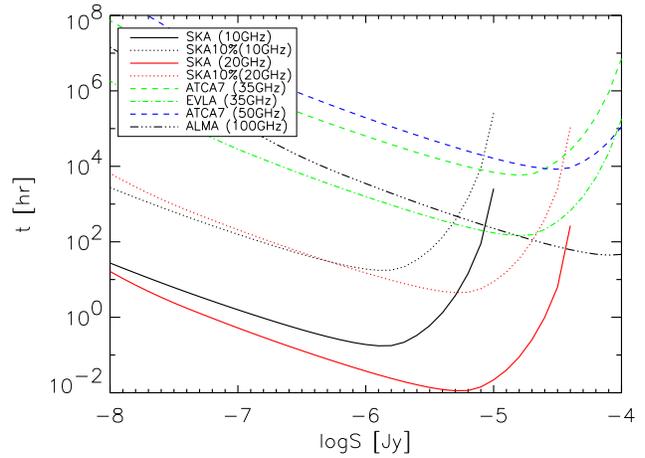}
\end{center}
\caption{Total survey time for ALMA, SKA, EVLA and ATCA to detect
100 protospheroids in thermal SZ at the frequencies specified in
the inset.} \label{fig:bestconf}
\end{figure}
\begin{figure}
\begin{center}
\includegraphics[width=6cm, angle=90]{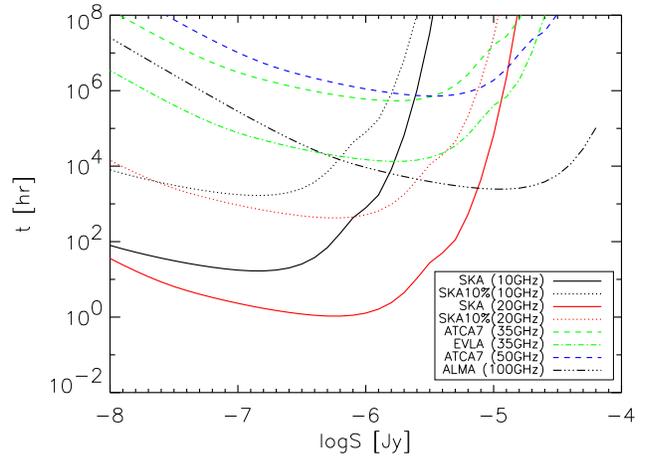}
\end{center}
\caption{Same as figure \ref{fig:bestconf} but for the kinetic SZ
effect.} \label{fig:bestconf_SZk}
\end{figure}

The time necessary to reach the wanted $S_{\rm lim}$ with $S/N=5$
in a single pointing, $t_{\rm p}$, is obtained from
eq.~(\ref{eq:sensitivity}), and the total observing time for
detecting $N_{\rm s}$ sources (excluding the slew time) is
obviously $t_{\rm p}n_{\rm p}$. In Figs.~\ref{fig:bestconf} and
\ref{fig:bestconf_SZk} we show the on-source time $t_{\rm p}n_{\rm
p}$ for $N_{\rm s}=100$ as a function of the absolute value of the
thermal and kinetic SZ limiting flux for the 4 instruments in
Table~\ref{tab:interferometers} at the frequencies specified in
the inset. The curves have minima at the values of $S_{\rm lim}$
corresponding to the fastest survey capable of detecting the
wanted number of sources. Clearly, it will be very time-consuming
to detect 100 protospheroids with the EVLA, and unrealistic with
the ATCA. On the other hand, since the 7 mm upgrade of ATCA will
be operational already in 2008, it will be possible to exploit it
to get the first test of the present predictions, and possibly to
achieve the first detection of an SZ signal from a
proto-spheroidal galaxy.

The SKA large effective collecting area allows the detection of thermal SZ
signals of 100 protospheroidal galaxies at 20(10) GHz in 1(11)
minutes with 7(7) pointings reaching $S_{\rm
lim}=10^{-5.3}(10^{-5.9})\,$Jy in a $0.46(0.47)\,\hbox{deg}^2$ area. The
10\% SKA requires  100 times more time than the full SKA but is
still faster than EVLA or ALMA. If phased array feeds were
available at the higher frequencies they would improve these
surveying times by a factor of up to 50.%

%
\subsection{Redshift distributions}
\begin{figure}
\begin{center}
\includegraphics[width=6cm, angle=90]{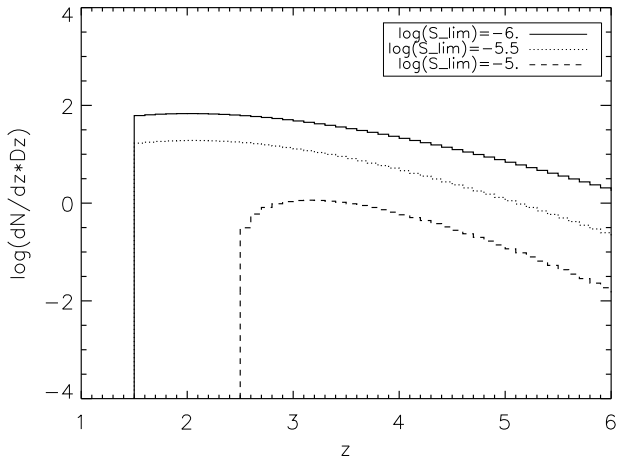}
\end{center}
\caption{Redshift distribution (in bins of width $\delta z= 0.1$) of
thermal Sunyaev-Zel'dovich effects at 20 GHz for the flux limits
specified in the inset. } \label{fig:comp_z_SZ}
\end{figure}
\begin{figure}
\begin{center}
\includegraphics[width=6cm, angle=90]{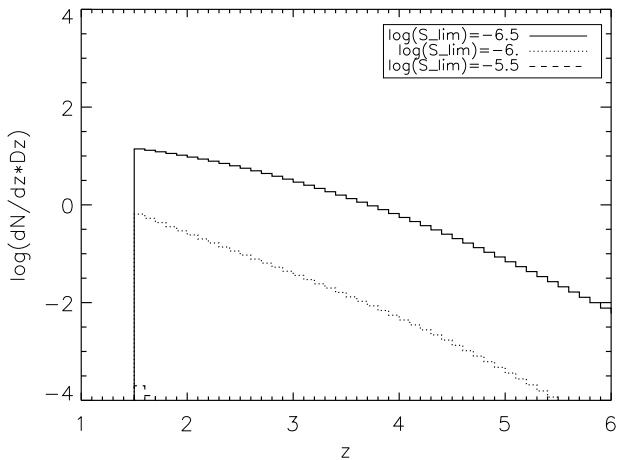}
\end{center}
\caption{Same as in Fig.~\protect{\ref{fig:comp_z_SZ}} but for the
kinetic Sunyaev-Zel'dovich effect.} \label{fig:comp_z_SZk}
\end{figure}
The redshift distributions of thermal and kinetic SZ effects are
illustrated, for 3 values of $S_{\rm lim}$, in
Figs.~\ref{fig:comp_z_SZ} and \ref{fig:comp_z_SZk}. They are both
relatively flat, as the fast decrease with increasing $z$ of the
density of massive (i.e. SZ bright) halos is partially compensated
by the brightening of SZ signals [eqs.~(\ref{eq:S_tSZ}) and
(\ref{eq:S_kSZ})]. Such brightening is stronger for the thermal
than for the kinetic SZ. A consequence of such brightening is that
the range of halo masses yielding signals above a given limit
shrinks with decreasing redshift, as the minimum detectable halo
mass increases. The upper limit on masses of galactic halos then
translates in a lower limit to the redshift distribution for
bright $S_{\rm lim}$.

\subsection{Contaminant emissions}\label{sec:contaminants}

The adopted model envisages that the plasma halo has the same size
as the dark matter halo, i.e. of order of hundreds kpc. In the
central region (with size of order of 10 kpc), the gas cools
rapidly and forms stars. Bressan, Silva, \& Granato (2002)
obtained a relationship between the star formation rate (SFR) and
the radio luminosity at 8.4 GHz, $L_S(8.4{\rm GHz})$:
\begin{equation} \label{eq:bressan}
L_S(8.4{\rm GHz}) \simeq 3.6\times 10^{27} {\hbox{SFR}\over
M_\odot/\hbox{yr}}\ \hbox{erg}\,\hbox{s}^{-1}\,\hbox{Hz}^{-1}.
\end{equation}
This relationship is in good agreement with the estimate by
Carilli (2001) while the equations in Condon (1992) imply a radio
luminosity about a factor of 2 lower, at fixed SFR.  The Granato
et al. (2004) model gives the SFR as a function of galactic age,
$t_{\rm gal}$, for any value of the the halo mass and of the
virialization redshift (see, e.g., Fig. 1 of Mao et al. 2007). We
must, however, take into account that eq.~(\ref{eq:bressan}) has
been derived using a Salpeter (1955) Initial Mass Function (IMF).
For the IMF used by Granato et al. (2004), the radio luminosity
associated to a given SFR is higher by a factor of 1.6 (Bressan,
personal communication). The coefficient in eq.~(\ref{eq:bressan})
was therefore increased by this factor.

The mean rest frame $8.4\,{\rm GHz}$ luminosity at given $M_{\rm
vir}$ and $z_{\rm vir}$ was then obtained using the corresponding
SFR averaged over $t_{\rm gal}$ in the renormalized
eq.~(\ref{eq:bressan}). Extrapolations in frequency have been
obtained using, as a template, the fit to the Arp~220 continuum
spectrum obtained by Bressan et al. (2002; solid line in their
Fig.~2). Using the continuum spectrum of M~82 (solid line in
Fig.~1 of Bressan et al.), the other standard starburst template,
we get essentially identical results.
\begin{figure}
\begin{center}
\includegraphics[width=6cm, angle=90]{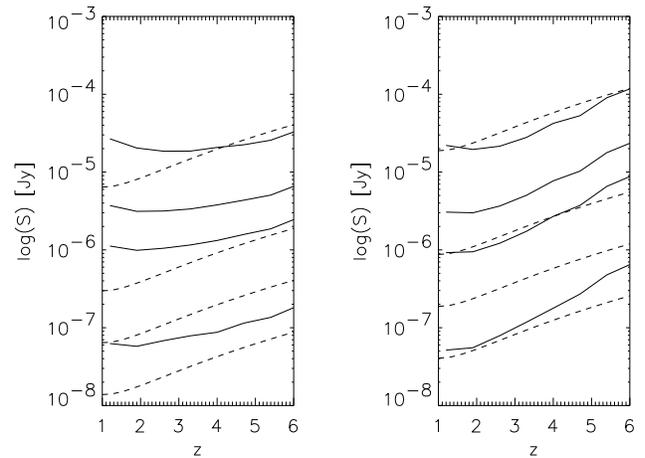}
\end{center}
\caption{Comparison of the flux associated with star formation
(solid lines) with the thermal SZ ``flux'' (dashed lines) at $20{\rm
GHz}$ (left panel) and at $35{\rm GHz}$ (right panel), as a function
of the virialization redshift for four values of the virial mass
($\log(M_{\rm vir})$= 11.5, 12., 12.5, 13.2, from bottom to top).}
\label{fig:contamination}
\end{figure}
In Fig. \ref{fig:contamination} we compare the flux associated
with star formation with the thermal SZ ``flux''
[eq.~(\ref{eq:S_tSZ})], as a function of the virialization
redshift, for several values of $M_{\rm vir}$. For a given halo
mass, the ratio of the thermal SZ to the contaminating signal
increases with frequency (and with redshift) as far as the
contamination is due to radio emission associated to star
formation (we do not consider here nuclear radio emission, which
occurs in $\lsim 10$ per cent of galaxies). However, already at 20
GHz the thermal dust emission becomes important for the highest
redshift sources. Such emission is more steeply increasing with
frequency than the SZ signal, even in the Rayleigh-Jeans region of
the CMB, and rapidly overwhelms it at $\gsim 100\,$GHz. The
SZ/contamination ratio increases with increasing halo mass;
therefore the SZ detection is easier for the more massive halos.
Thus in the range 10--35 GHz the thermal SZ is expected to
dominate over the contaminating signal at least for the most
massive objects.

It must be noted that the star forming regions are concentrated in
the core of the spheroids, on angular scales of the order or less
than $1''$, for the redshifts considered here. Long ($\gsim 3\,
\rm km$ at $35\, \rm GHz$ for full sensitivity) baselines
observation with high sensitivity may be able to resolve the star
forming region positive signal and subtract it from the image. To
achieve this purpose a good sampling of the shortest spacings on
the uv plane is necessary together with a good sampling of the
largest ones: with the latter it might be possible to reconstruct
the contaminated profile, subtract it from the former and produce
an uncontaminated SZ profile. Again, the SKA at high frequencies
seems to be the optimal instrument.

\subsection{Confusion effects}\label{sec:confusion}

Further constraints to the detection of SZ effects are set by
confusion fluctuations. Fomalont et al. (2002) have determined the
8.4 GHz source counts down to $7.5\,\mu$Jy. For $S_{8.4{\rm GHz}}
\lsim 1\,$mJy they are well described by:
\begin{equation}\label{eq:Fomalont}
N(>S)\simeq 1.65\times 10^{-3} S^{-1.11}\ \hbox{arcsec}^{-2}
\end{equation}
with $S$ in $\mu$Jy. The spectral index distribution peaks at
$\alpha \simeq 0.75$ ($S \propto \nu^{-\alpha}$).

For all but one (SKA 10 GHz) of the considered surveys the
``optimal'' depth for detecting 100 sources corresponds to 8.4 GHz
flux densities within the range covered by Fomalont et al. (2002),
so that the confusion fluctuations are dominated by sources obeying
eq.~(\ref{eq:Fomalont}). We then have:
\begin{equation}\label{eq:conf}
\sigma^2_{\rm conf}\simeq 0.2\left({\nu \over 8\ {\rm
GHz}}\right)^{-1.11\alpha} {\omega\over 100\ {\rm arcsec}^2}S_{\rm
d}^{0.89}\ \mu\hbox{Jy}^2
\end{equation}
where $S_{\rm d}$, in $\mu$Jy, is the detection limit and $\omega$
is the solid angle subtended by the SZ signal.
Equation~(\ref{eq:conf}) can be rewritten as
\begin{equation}\label{eq:conf2}
{S_{\rm d}\over \sigma_{\rm conf}}\simeq 2.2\left({\nu \over 8.4\
{\rm GHz}}\right)^{0.555\alpha} \left({\omega\over 100\ {\rm
arcsec}^2}\right)^{-1/2}S_{{\rm d},\mu{\rm Jy}}^{0.555},
\end{equation}
yielding a $5\sigma_{\rm conf}$ detection limit of $\simeq 4\,\mu$Jy
at 10 GHz and of $\simeq 2.3\,\mu$Jy at 20 GHz. For the ``optimal''
survey depths at higher frequencies $S_{\rm d}/\sigma_{\rm conf}\gg
5$, implying that they are not affected by confusion noise due to
radio sources.

On the other hand, as noted above, at high frequencies the
redshifted dust emission from distant star-forming galaxies
becomes increasingly important (De Zotti et al. 2005). To estimate
their contribution to the confusion noise, we have used once again
the model by Granato et al. (2004), with the dust emission spectra
revised to yield $850\mu$ counts consistent with the results by
Coppin et al. (2006), and complemented by the phenomenological
estimates by Silva et al. (2004, 2005) of the counts of sources
other than high-$z$ proto-spheroids (see Negrello et al. (2007)
for further details). We find, for a typical solid angle $\omega =
100\,{\rm arcsec}^2$, $5\sigma_{\rm conf}$ flux limits due to
these sources of 3, 55, and $190\,\mu$Jy at 20, 35, and 50 GHz,
respectively. Thus at 20 GHz we have significant contributions to
the confusion noise both from the radio and from the dust
emission; the overall $5\sigma_{\rm conf}$ detection limit is
$S_{\rm d} \simeq 4\,\mu$Jy. At 10 GHz the contribution of dusty
galaxies to the confusion noise is negligible, while at 100 GHz
the confusion limit is as high as 2 mJy, implying that the
detection of the galactic-scale SZ effect is hopeless at mm
wavelengths.

Although high-$z$ luminous star-forming galaxies may be highly
clustered (Blain et al. 2005; Farrah et al. 2006; Magliocchetti et
al. 2007), the clustering contribution to fluctuations is negligible
on the small scales of interest here (De Zotti et al. 1996), and can
safely be neglected.

\section{Conclusions}\label{sec:conclusions}

In the standard scenario for galaxy formation, the proto-galactic
gas is shock heated to the virial temperature. The observational
evidences that massive star formation activity must await the
collapse of large halos, a phenomenon referred to as {\it
downsizing}, suggest that proto-galaxies with a high thermal energy
content existed at high redshifts. Such objects are potentially
observable through the thermal and kinetic Sunyaev-Zel'dovich
effects and their free-free emission. The detection of this phase of
galaxy evolution would shed light on the physical processes that
govern the collapse of primordial density perturbations on galactic
scales and on the history of the baryon content of galaxies.

As for the latter issue, the standard scenario, adopted here,
envisages that the baryon to dark matter mass ratio at virialization
has the cosmic value, i.e. is about an order of magnitude higher
than in present day galaxies. Measurements of the SZ effect will
provide a direct test of this as yet unproven assumption, and will
constrain the epoch when most of the initial baryons are swept out
of the galaxies.

As mentioned in \S\,\ref{sec:intro},  almost all semi-analytic
models for galaxy formation adopt halo mass functions directly
derived or broadly consistent with the results of N-body
simulations, and it is commonly assumed that the gas is shock
heated to the virial temperature of the halo. They therefore
entail predictions on counts of SZ effects similar to those
presented here. On the other hand, the thermal history of the gas
is governed by a complex interplay of many astrophysical
processes, including gas cooling, star formation, feedback from
supernovae and active nuclei, shocks. As mentioned in
\S\,\ref{sec:intro}, recent investigations have highlighted that a
substantial fraction of the gas in galaxies may not be heated to
the virial temperature. Also, it is plausible that the AGN
feedback transiently heats the gas to temperatures substantially
above the virial value, thus yielding SZ signals exceeding those
considered here. The gas thermal history may therefore be
substantially different from that envisaged by semi-analytic
models, and the SZ observations may provide unique information on
it.

We have presented a quantitative investigation of the counts of SZ
and free-free signals in the framework of the Granato et al.
(2004) model, that successfully accounts for the wealth of data on
the cosmological evolution of spheroidal galaxies and of AGNs
(Granato et al. 2004; Cirasuolo et al. 2005; Silva et al. 2005;
Lapi et al. 2006).

We find that the detection of substantial numbers of galaxy-scale
thermal SZ signals is achievable by blind surveys with next
generation radio interferometers. Since the protogalaxy thermal
energy content increases, for given halo mass, with the
virialization redshift, the SZ ``fluxes'' increase rather strongly
with $z$, especially for the thermal SZ effect, partially
compensating for the rapid decrease of the density of massive halos
with increasing redshift. The redshift distributions of thermal SZ
sources are thus expected to have substantial tails up to high $z$.

There are however important observational constraints that need to
be taken into account. The contamination by radio and dust
emissions associated to the star formation activity depends on
mass and redshift of the objects, but is expected to be stronger
than the SZ signal at very low and very high frequencies. We
conclude that the optimal frequency range for detecting the SZ
signal is from 10 to 35 GHz, where such signal dominates over the
contamination at least for the most massive objects. It must be
noted however that contaminating emissions have typical scales of
the order of those of the stellar distributions, i.e. $< 1''$ at
the redshifts of interest here (see Fomalont et al. 2006), while
the SZ effects show up on the scale of the dark matter halo, which
is typically ten times larger. Therefore arcsec resolution images,
such as those that will be provided by the SKA, will allow to
reconstruct the uncontaminated SZ signal.

The coexistence of the hot plasma halo, responsible for the SZ
signal, with dust emission implies that the scenario presented in
this paper may be tested by means of pointed observations of
high-$z$ luminous star-forming galaxies detected by (sub)-mm
surveys.

Confusion noise is a very serious limiting factor at mm
wavelengths. Contributions to confusion come on one side from
radio sources and on the other side from dusty galaxies. At 10 GHz
only radio sources matter; a modest extrapolation of the 8.4 GHz
$\mu$Jy counts by Fomalont et al. (2002) gives a $5\sigma_{\rm
conf}$ detection limit $S_{\rm d} \simeq 4\,\mu$Jy, for a SZ
signal subtending a typical solid angle of $100\,\hbox{arcsec}^2$.
Fluctuations due to dust emission from high-$z$ luminous
star-forming galaxies may start becoming important already at 20
GHz; at this frequency, quadratically summing them with those due
to radio sources we find again $S_{\rm d} \simeq 4\,\mu$Jy, for
the same solid angle. On the other hand, the high resolution of
the SKA and of the EVLA will allow us to effectively detect and
subtract out confusing sources, thus substantially decreasing the
confusion effects. Beating confusion will be particularly
important for searches of the weaker kinetic SZ signal.

\section*{ACKNOWLEDGMENTS}
We are indebted to G.~L. Granato for having provided the software
to compute the cooling function and the halo mass function, to A.
Diaferio for explanations on the halo velocity function and its
cosmological evolution, and to A. Bressan for clarifications on
the dependence of the radio-luminosity/SFR relation on the IMF.\\
We thank M. Rupen for updated information about the EVLA capabilities.\\
We also warmly thank the anonymous referee for an unusually
accurate reading of the manuscript and for many constructive
comments.\\
Work supported in part by the Ministero dell'Universit\`a e della
Ricerca and by ASI (contract Planck LFI Activity of Phase E2).

\end{document}